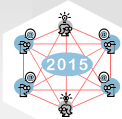

# OMNeT++ and mosaik: Enabling Simulation of Smart Grid Communications
## Position Paper

Jens Dede*, Koojana Kuladinithi*, Anna Förster*, Okko Nannen†, Sebastian Lehnhoff†
*Sustainable Communication Networks Group, University of Bremen, Germany
Email: {jd, koo, afoerster}@comnets.uni-bremen.de
†R&D Division Energy, OFFIS, Germany
Email: {okko.nannen, sebastian.lehnhoff}@offis.de

*Abstract*—This paper presents a preliminary system architecture of integrating OMNeT++ into the mosaik co-simulation framework. This will enable realistic simulation of communication network protocols and services for smart grid scenarios and on the other side, further development of communication protocols for smart grid applications. Thus, by integrating OMNeT++ and mosaik, both communities will be able to leverage each others's sophisticated simulation models and expertise.

The main challenges identified are the external management of the OMNeT++ simulation kernel and performance issues when federating various simulators, including OMNeT++ into the mosaik framework. The purpose of this paper is to bring these challenges up and to gather relevant experience and expertise from the OMNeT++ community. We especially encourage collaboration among all OMNeT++ developers and users.

## I. INTRODUCTION

Smart grids have been identified as the most important development towards renewable energy production and usage. Still in their baby shoes, they exhibit a number of challenges in a very wide spectrum, reaching from market strategies through new power grid networks to communication.

Mosaik [1] is a flexible smart grid co-simulation framework. It allows to combine several simulators for different requirements. Using mosaik, it is possible to combine simulators for photovoltaic, power plants, households etc. to a complex smart grid simulation scenario. However, mosaik assumes a perfect link for the communication between the individual entities like houses, power plants and electric vehicles.

OMNeT++ focusses on the communication link and offers a huge variety of toolboxes for the majority of common communication technologies.

The purpose of this position paper shows a possible way of connecting OMNeT++ and mosaik to get an easy-to-use framework for a realistic simulation of the communication and the power flows of current and future smart grid scenarios.

## II. RELATED WORKS

Several publications are available which introduce co-simulation solutions for the power grid and communication links. Most of the work highlight the integration of the different simulator types as the main issue: Most power grid simulators operate using discrete time simulation whereas communication network simulators discretise on an event base (discrete event simulation). One task for co-simulation frameworks is to combine both simulator types. Without claiming completeness, we point out some exemplary existing solutions.

The authors in [2] simulate the influence of cyber-attacks on a Smart Grid using Matlab/Simulink in combination with OPNET. Due to the change of the license policy of riverbed (manufacturer of OPNET) for academic users, OPNET has lost however its attractiveness for universities. In [3], a generic co-simulation framework called "*FNCS*" is introduced. It can connect power grid and communication network simulators. The evaluation of the framework is performed using GridLAB-D (distribution simulator), PowerFlow (power flow simulator) and ns-3 (communication simulator). It can be extended to use other simulators. Additional time synchronisation algorithms for FNCS were introduced in [4]. Our own extensive work on communication protocols focuses on OMNeT++ as a tool and thus transferring is not feasible. Anyway, some possible solutions to overcome co-simulation issues regarding discrete time and discrete event simulation are discussed in FNCS publications and might be helpful for the desired extension of OMNeT++.

Also for OMNeT++, co-simulation frameworks for the smart grid are available. In [5], OMNeT++ and OpenDSS (electric power Distribution System Simulator) are combined and OMNeT++ takes control of OpenDSS. In [6], OMNeT++ and OpenDSS are run in parallel and the events are synchronized at certain time slots. Both solutions are limited to OpenDSS for the grid simulation and not easily extendible for additional simulators.

In [7], the authors use OMNeT++ to analyze measurements from a real testbed to evaluate the communication effort caused by using electric vehicles for stabilizing the power grid. This shows a good example of the flexibility of OMNeT++ regarding combination with other information sources.

In summary, the existing co-simulation frameworks support only few power grid simulators like OpenDSS, GridLAB-D and PowerFlow. Mosaik offers a framework with support for a considerable number of simulators. Plus, it offers simulation servers and support. Combining OMNeT++ and mosaik will result in a flexible and powerful co-simulation framework not only for the power aspects but also for the required communication networks of future power grids. Furthermore, this integration will utilise a variety of communication technologies and protocols already available on OMNeT++ to be evaluated for future applications in the area of power grids.

### III. OVERVIEW OF MOSAIK

Smart grids comprise a vast number of active components and participants that have to fulfil the task of keeping power demand and supply in balance while adhering to delicate operational constraints, which are required to maintain a safe and a stable power supply. Further, the domain space of future energy systems, so called "Smart Grids" are closely interlinked with other complex systems as given below.

- environmental weather conditions (influencing wind and solar power feed-in)
- energy markets (influencing unit commitment and large-scale generation schedules)
- socio-technical system (influencing end-user-operated appliances as well as small scale generation units)
- information and communication systems (influencing the availability and accuracy of operational telemetry data necessary for a stable and efficient operation)

Sophisticated models and simulators exist for each of the above mentioned systems that have been researched and expensively developed by respective domain experts. Though some of the existing smart grid simulation tools have extended their scope to consider other relevant systems (e.g. power system plus communication system simulation), these setups are usually limited in functionality of the whole spectrum of future energy domain. To the best of our knowledge, no simulators yet exist that are capable of producing the entire domain spectrum for smart grid applications [8]. This is simply because of complexity in integrating above mentioned systems developed individually to model more specific functions. However, several existing co-simulation suites exist as mentioned in the related work.

The SESA-Lab at the University of Oldenburg aims at bridging this gap by providing a collaborative simulation and integration platform that consists of sophisticated adapters/interfaces to many interdisciplinary tools and simulators in the domain space of future energy systems. It is continuously extended to functionally interlink domain-specific models that have not yet been able or used to run in an integrated fashion with other sophisticated simulators. The proposed simulation framework is called mosaik that not only provides interfaces to the above mentioned systems, but a scheduling mechanism that is capable of efficiently orchestrating the communication and information exchange between heterogeneous models and simulators (hard- and software) interoperated in this fashion. Mosaik allows you to reuse and combine existing simulation models and simulators to create large-scale Smart Grid scenarios – i.e. thousands of simulated entities distributed over multiple simulator processes. These scenarios can then serve as a test bed for various types of control strategies (e.g., multi-agent systems (MAS) or centralized control). This is shown in Figure 1 which presents the overall system architecture of mosaik with several connected simulators.

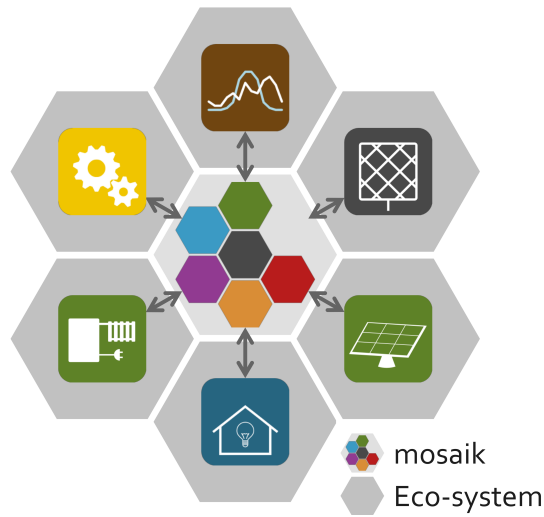

Fig. 1. The mosaik system architecture: mosaik orchestrates all connected simulators

Mosaik assumes a perfect communication link between the separate simulated entities. To overcome this limitation, we will extend OMNeT++ with mosaik interfaces to simulate realistic links. Mosaik is written in Python and completely open source (under LGPL v2.0), including some simple simulators, a binding to PYPOWER and a demonstration scenario. More information can be found at [1].

### IV. OVERVIEW OF OMNET++

This paper focuses on how to enable the integration of OMNeT++ into the mosaik environment. We will not go into details about the OMNeT++ simulation environment here, since we assume a high degree of expertise in the audience. OMNeT++ is a discrete event simulator for simulating communication/interaction between some network entities or agents. Typically these agents are fully controlled by OMNeT++, even if their implementation might be (partially) external, e.g. through additional libraries. The main reason of doing this is the nature of discrete event simulators, where events are inserted from all communicating agents into one single event queue, which is processed then sorted in terms of simulation time.

Currently, there are possibilities to embed OMNeT++ simulations into other applications, which means in particular that the simulation can be started externally and results can be fed into the calling application automatically. Furthermore, as already mentioned above, individual agents can be implemented externally, but are controlled from inside the OMNeT++ simulation. For example, one can use an implementation of a battery behaviour from an external library, which is called from inside a corresponding OMNeT++ module. This makes it very easy to integrate simulations **into** OMNet++, but not the other way around. There are no possibilities to control OMNeT++ externally, e.g. calling it event by event or inserting events from other applications. At the same time, the highly modularised architecture of OMNeT++ makes this more an implementation question rather than a technical limitation or real challenge.

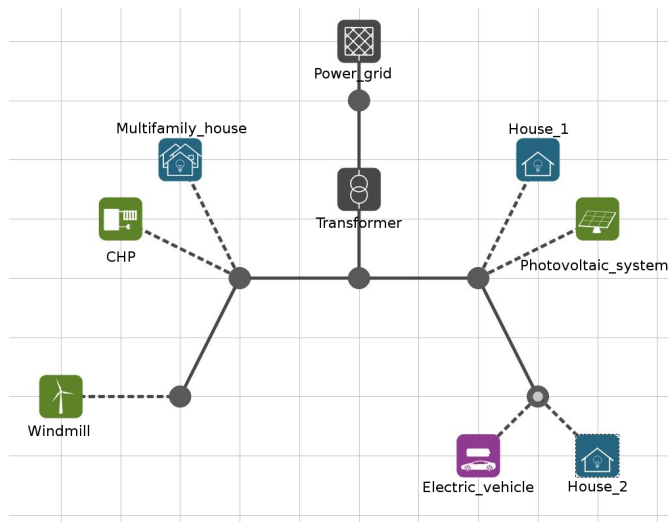

(a) The mosaik representation

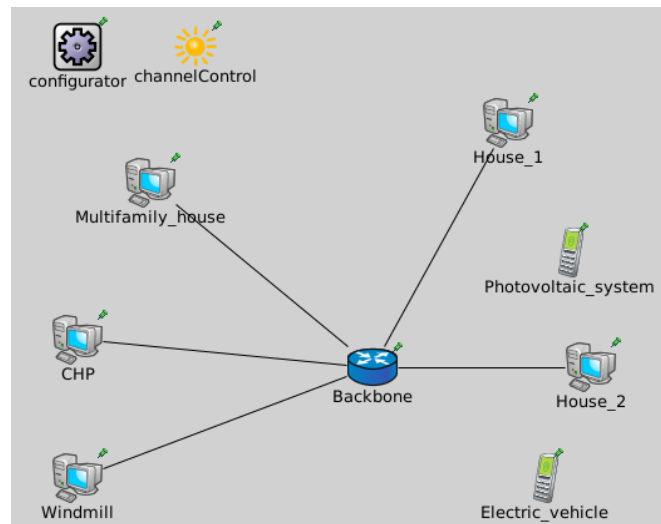

(b) The OMNeT++ representation

Fig. 2. Two representations of one scenario containing a windmill-powered plant, a combined heat and power unit (CHP), one multifamily house, a transformer, an uplink to the upper layer power grid as a reference, two houses, one photovoltaic system and an electric vehicle. (a) depicts the corresponding mosaik representation while (b) shows the same in OMNeT++. Note: The transformer and the upper layer power grid are not depicted as they are assumed to not have communication capabilities.

## V. Preliminary System Architecture

The system architecture of mosaik is depicted in Figure 1. The communication between simulators is enabled in a direct way through the simulation kernel itself. This means, if two different entities in a simulation need to exchange information, they do so by simply sending the information to the other entity through mosaik. No delays, protocol details, etc. are considered. The corresponding API was published first in [9].

In order to adhere to the general system architecture of mosaik, OMNeT++ needs to be connected as a mosaik component through implementing the mosaik API. Then, communication between entities will be forwarded to OMNeT++ for simulation instead of directly forwarding to the receiver entity. Besides the implementation and integration of the mosaik API, the time synchronization of mosaik and OMNeT++ has to be considered. While OMNeT++ performs a discrete event simulation, mosaik performs a discrete time simulation. The synchronization of both types of simulation is for example discussed in [3] and [4]. Besides the time synchronization, several additional adaptations are required:

### A. Extracting the relevant communication network from mosaik to OMNeT++

Figure 2 depicts an example scenario consisting of a windmill-powered plant, a combined heat and power unit (CHP), one multifamily house, a transformer, an uplink to the upper layer power grid as a reference, two houses, one photovoltaic system and an electric vehicle. The power electric connections of all entities are show in the topology representation of mosaik as depicted in Figure 2(a). Mosaik does not consider the communication links. Figure 2(b) shows a possible representation of the communication links using OMNeT++ where we assume that the transformer and the upper layer power grid have no communication capabilities and therefore are not shown. Furthermore, the electric vehicle and the photovoltaic system are assumed to communicate wirelessly as entities like the photovoltaic system might be placed in a remote place without wired connectivity. An electric vehicle may require a wireless connection to request load capacities, i.e. while driving home.

To co-simulate the communication using mosaik, the communication capabilities of all entities in mosaik are required and need to be transfered to OMNeT++. We consider this step is only moderately challenging, as it only includes extensions to mosaik, which fully conform to its current design. From the side of OMNeT++, a NED description can be extracted relatively easy from the mosaik scenario description and thus easily forwarded to OMNeT++.

### B. Full external control of OMNeT++ simulation

In order to give the full control over the simulation to mosaik to orchestrate all simulators at the same time, OMNeT++ needs to be controlled externally. Currently, there is no obvious way to do this. The only relevant feature is the possibility to embed the simulation, but this does not give the control to an external party. Rather, we need to do the following:

- First initialise the simulation
- Run events one by one and pass relevant results to mosaik
- Insert new events externally from mosaik (e.g. a new data packet or stream)
- Forward errors to mosaik
- Finish the simulation

As mentioned above, the exchange of events between mosaik and OMNeT++ requires a synchronization between the mosaik time base and the OMNeT++ event base. We plan to implement this by extending the OMNeT++ kernel with methods similar to the already existing `simulate()` function, such as `simulate_with_control()`, `step_one_event()` and `simulate_until()`.

Besides the full external control of OMNeT++, a real-time simulation of both simulators might be feasible. In this case, mosaik could insert events at the application layer to the OMNeT++ simulator. But this may lead to the following issues:

- As the runtime of targeted smart grid simulations can be long (i.e. in the order of month or even years), the simulation will be long as well.
- For complex simulation models and scenarios, the real-time simulation might be not applicable due to the limitation of processing power (simulation time longer than real time).

*C. Improve performance of federated simulations*

The above described idea of running OMNeT++ externally from mosaik event by event is not very efficient. It will require mosaik to call OMNeT++ quite often, even if no relevant events are executed. For example, if two entities are trying to communicate through OMNeT++ using a UDP connection, the only relevant events for the entities are the passing of the data to the connection and the received data on the other side. All protocol-level events, such as queueing, medium access, etc. do not have to be exposed to mosaik. Therefore, several options can be used to optimise the performance:

- Pass results to mosaik only at the OMNeT++ application level, e.g. as defined in the INET framework. All other events are kept internal and are not communicated to mosaik. This results in the issues mentioned in Section V-B.
- Enable multiple event simulation in OMNeT++. The main challenge lies in the fact that OMNeT++ does not know when the next communication is due, as this is in control of mosaik. Thus, mosaik needs to resolve in its kernel when the next communication is due and to allow OMNeT++ to run so far.
- Identify independent communications. Mosaik needs to explore its scenario and to identify communication islands. In this case, several smaller OMNeT++ simulations can be started and executed in parallel. But it is a challenging task to explore these communication island in a reliable way.
- Make use of existing parallelization techniques for OMNeT++. This will speed up OMNeT++ itself and thus also the general mosaik scenario.

## VI. CONCLUSION

In this position paper, the integration of OMNeT++ into the mosaik co-simulation framework was described. Several pitfalls and required adaptations were brought up. However, no serious limitations or expected problems have been identified and we believe that the federation of the testbeds is feasible. The main issue thus remains the performance of the federated simulations.

## REFERENCES

[1] "Mosaik is a flexible smart grid co-simulation framework." accessed: 2015-06-23. [Online]. Available: http://mosaik.offis.de

[2] M. A. H. Sadi, M. H. Ali, D. Dasgupta, and R. K. Abercrombie, "Opnet/simulink based testbed for disturbance detection in the smart grid," in *Proceedings of the 10th Annual Cyber and Information Security Research Conference*, ser. CISR '15. New York, NY, USA: ACM, 2015, pp. 17:1–17:4. [Online]. Available: http://doi.acm.org/10.1145/2746266.2746283

[3] S. Ciraci, J. Daily, J. Fuller, A. Fisher, L. Marinovici, and K. Agarwal, "Fncs: A framework for power system and communication networks co-simulation," in *Proceedings of the Symposium on Theory of Modeling & Simulation - DEVS Integrative*, ser. DEVS '14. San Diego, CA, USA: Society for Computer Simulation International, 2014, pp. 36:1–36:8. [Online]. Available: http://dl.acm.org/citation.cfm?id=2665008.2665044

[4] S. Ciraci, J. Daily, K. Agarwal, J. Fuller, L. Marinovici, and A. Fisher, "Synchronization algorithms for co-simulation of power grid and communication networks," in *Modelling, Analysis Simulation of Computer and Telecommunication Systems (MASCOTS), 2014 IEEE 22nd International Symposium on*, Sept 2014, pp. 355–364.

[5] M. Lévesque, D. Q. Xu, G. Joós, and M. Maier, "Communications and power distribution network co-simulation for multidisciplinary smart grid experimentations," in *Proceedings of the 45th Annual Simulation Symposium*, ser. ANSS '12. San Diego, CA, USA: Society for Computer Simulation International, 2012, pp. 2:1–2:7. [Online]. Available: http://dl.acm.org/citation.cfm?id=2331751.2331753

[6] D. Bhor, K. Angappan, and K. Sivalingam, "A co-simulation framework for smart grid wide-area monitoring networks," in *Communication Systems and Networks (COMSNETS), 2014 Sixth International Conference on*, Jan 2014, pp. 1–8.

[7] S. Bocker, C. Lewandowski, C. Wietfeld, T. Schluter, and C. Rehtanz, "Ict based performance evaluation of control reserve provision using electric vehicles," in *Innovative Smart Grid Technologies Conference Europe (ISGT-Europe), 2014 IEEE PES*, Oct 2014, pp. 1–6.

[8] M. Buscher, A. Claassen, M. Kube, S. Lehnhoff, K. Piech, S. Rohjans, S. Scherfke, C. Steinbrink, J. Velasquez, F. Tempez, and Y. Bouzid, "Integrated smart grid simulations for generic automation architectures with rt-lab and mosaik," in *Smart Grid Communications (SmartGridComm), 2014 IEEE International Conference on*, Nov 2014, pp. 194–199.

[9] S. Schütte, S. Scherfke, and M. Sonnenschein, "mosaik-smart grid simulation api," *Proceedings of SMARTGREENS*, pp. 14–24, 2012.